\documentclass[11pt,preprint,showpacs,preprintnumbers,amsmath,amssymb]{revtex4}
\usepackage{amsmath}
\usepackage{graphicx}
\usepackage{amssymb}
\usepackage{graphicx}% Include figure files
\usepackage{dcolumn}% Align table columns on decimal point
\usepackage{bm}% bold math

\begin{document}

\preprint{}

\title{Exact asymptotics for non-radiative migration-accelerated energy transfer
in one-dimensional systems.}% Force line breaks with \\

\author{G.Oshanin}
\email{oshanin@lptmc.jussieu.fr}
 \affiliation{Laboratoire
de Physique Th\'eorique de la Mati\`ere Condens\'ee (CNRS-UMR 7600), Universit\'e
Pierre et Marie Curie,
Tour 24, Boite 121, 4 place Jussieu, F-75252 Paris 05, France}%Lines break automatically or can be forced with \\
\author{M.Tachiya}%
 \email{m.tachiya@aist.go.jp}
\affiliation{National Institute of Advanced Industrial Science and
Technology (AIST), 1-1-1 Higashi, Tsukuba, Ibaraki 305-8565, Japan}%

\date{\today}% It is always \today, today,

\begin{abstract}
We study direct energy transfer by multipolar  or exchange
interactions between diffusive excited donor and diffusive unexcited
acceptors. Extending over the case of long-range transfer of an
excitation energy a non-perturbative approach by Bray and Blythe
[Phys. Rev. Lett. {\bf 89}, 150601 (2002)], originally developed for
contact diffusion-controlled reactions, we determine exactly
long-time asymptotics of the donor decay function in one-dimensional
systems. 
\end{abstract}

\maketitle

\section{Introduction}

Long-range non-radiative transfer of an excitation energy  from
excited donor molecules to acceptors of the excitation energy  is a
dominant reaction mechanism in various chemical, physical and
biological processes \cite{rice,bur,graetzel,klafter,bar}. To name
but a few, we mention fluorescence, luminescence or phosphorescence
quenching, decay of trapped electrons in glassy media in presence of
scavengers, or light harvesting by antennae chlorophyll-b molecules
and donation of singlet energy to the chlorophyll-a reaction centers
in photosynthetic organisms.

The idea of direct non-radiative transfer has been put forward in
the pioneering works of  F\"orster \cite{f} and Dexter \cite{d}, who
determined decay of an immobile excited donor due to dipole-dipole
interactions with immobile, randomly placed acceptors in rigid
three-dimensional solutions. Subsequent analysis (see Refs.\cite{2} and references therein) extended
the consideration of F\"orster and Dexter to arbitrary Euclidean
dimensions $d$ and to  general forms of donor-acceptor interactions,
such as isotropic multipolar interactions, for which the rate $W(r)$
of energy transfer is given by
\begin{equation}
\label{mult}  W(r) = \alpha_m \left(\frac{r_0}{r}\right)^n,
\end{equation}
or interactions mediated by exchange, for which one has
\begin{equation}
\label{exchange} W(r) = \alpha_e \exp\left(- \gamma r\right),
\end{equation}
where $r$ is the distance separating a given donor-acceptor pair,
the constants $n$, $r_0$ and $\gamma$ determine the interaction type
and range (e.g., $n=6$ for dipolar, $n=10$ for quadrupolar
interactions). 

Note that an exponential form in Eq.(\ref{exchange})
emerges, as well, in another important area - outer sphere electron
transfer reactions. Kinetics of such electron tunneling processes
taking place in liquids or glassy media have been also widely
studied giving rise to a very beneficial crossfertilization of ideas
and approaches.

For the transfer rates in Eqs.(\ref{mult}) and (\ref{exchange}), it
was found \cite{2} that the probability $P(t)$ that the donor is
still in excited state up to time $t$ obeys, at sufficiently long
times,
\begin{equation}
\label{mult1} P(t) \sim \exp\left[- V_d \Gamma(1-d/n) n_A r_0^d
\left(\alpha_m t\right)^{d/n}\right]
\end{equation}
for multipolar and
\begin{equation}
\label{exchange1} P(t) \sim \exp\left[- V_d \gamma^{-d} n_A
\ln^d\left(\alpha_e t\right)\right]
\end{equation}
for exchange-mediated transfer, respectively. In Eqs.(\ref{mult1})
and (\ref{exchange1}), $n_A$ denotes mean density of acceptor
molecules, $V_d = \pi^{d/2}/\Gamma(1+d/2)$ and $\Gamma(x)$ is a
Gamma-function. The decay forms in Eqs.(\ref{mult1}) and
(\ref{exchange1}) have been also generalized for certain types of
restricted geometries - fractals \cite{26}, porous
\cite{klafter,26,lev,lev2} and various microheterogeneous
\cite{microhet} media as well as polymer solutions
\cite{polexp,polteor}.

However, in many situations the donors and acceptors are not immobile. In
liquids, both donor and acceptor molecules perform diffusive motion.
In solids, excitations become delocalized because of incoherent
hopping between donor sites, which ultimately results in a diffusive
transport, although the decay kinetics may be still different from
that predicted for conventional diffusive motion - there 
always exists a finite probability that an
excitation remains on an intitially excited donor \cite{b,fed}.

 It
was recognized \cite{yt} that random migration of donor and acceptor
molecules leads to a much more efficient deactivation than the
direct transfer between the immobile species; in three-dimensions,
in particular, one finds that $P(t)$ obeys \cite{yt}
\begin{equation}
\label{3D} P(t) \sim \exp\left[- 4 \pi \left(D_A + D_D\right) R_{\rm
eff} n_A t\right],
\end{equation}
where $D_D$ and $D_A$ are donor and acceptor diffusion coefficients
and $R_{\rm eff}$ is the effective reaction radius. 

Note that the
result in Eq.(\ref{3D}) has been first obtained  for $W(\rho)$ in
Eq.(\ref{mult}) with $n=6$ and $D_A=0$ in Ref.\cite{pdg}, which
analyzed the relaxation of the nuclear magnetization in the presence
of paramagnetic impurities. Recently, the result in Eq.(\ref{3D})
has been generalized to 3D systems with donors and acceptors
performing anomalous, "fractional" diffusion \cite{seki-san}.

To the best of our knowledge, there do not exist analogous results 
for low-dimensional, (i.e. 1D and 2D)
systems, although the extension of
the approach of Refs.\cite{yt} 
for 1D and 2D situations seems to be straightforward. 
In doing so, one will find 
\begin{equation}
\label{low} P(t) \sim \exp\left[- n_A \phi(t) \right],
\end{equation}
where $\phi(t) \sim ((D_A + D_D) t)^{1/2}$ in 1D and $\phi(t) \sim
(D_A + D_D) t/\ln((D_A + D_D) t)$ in 2D.

On the other hand, it is clear that neither Eq.(\ref{3D}) nor Eq.(\ref{low}) is 
an exact solution
for the long-range energy transfer problem involving diffusive donor
and acceptor molecules, but a result of a certain assumption. While
particles' diffusion coefficients will indeed appear only in the
form of a sum $D_A + D_D$ in the solution of a problem with a single
donor and a single acceptor, it is not the case in the general
situation with a concentration of acceptor molecules. In particular,
setting $D_A = 0$, one should obtain a crossover to a singular
behavior characterized by a stretched-exponential time dependence
$\ln P(t) \sim - t^{d/(d+2)}$ \cite{pastur}, whereas Eq.(\ref{3D})
does not show any singularity in the limit $D_A \to 0$. 

Apart of this, it was
recently discovered \cite{bb}  that, remarkably, for diffusion-controlled 
contact $C + B \to B$ reactions taking place 
in low dimensional
systems (or, generally, in dimension $d \leq d_f$, where $d_f$ is
fractal dimension of particles' trajectories in case of subdiffusive
motion \cite{subdiff}), the long-time asymptotical form of $P(t)$ is
\textit{independent} of the $C$ particle diffusion coefficient.

In this paper we extend the non-perturbative approach of Bray and
Blythe \cite{bb} developed originally for contact
diffusion-controlled reactions to systems with diffusive donor and
acceptors interacting via distance-dependent isotropic multipolar or
exchange transfer rates in Eqs.(\ref{mult}) and (\ref{exchange}). We
define, in form of convergent in the limit $t \to \infty$ upper and
lower bounds, the exact form of the excitation survival probability
$P(t)$ in one-dimensional systems. 
More specifically, we
show that in 1D systems, both for multipolar and exchange-mediated
transfer $P(t)$ obeys
\begin{equation}
\label{1d1d} 1 + O\left(\frac{1}{t^{1/2}}\right) \leq \frac{\ln
P(t)}{- 4 n_A \sqrt{D_A t/\pi}} \leq 1 + O\left(
\frac{1}{t^{1/6}}\right).
\end{equation}
This exact result
proves that, in contrast to predictions based on standard
considerations, Eqs.(\ref{low}), the donor decay function in 1D is
independent of donor's diffusion coefficient in the asymptotic
regime. Remarkably, the decay forms appear to be \textit{exactly}
the same as for \textit{contact} 
diffusion-controlled trapping reactions, (such that reaction takes place upon
 encounters between particles), \cite{bb} despite the fact that in our case reaction
proceeds via \textit{long-range} transfer rates in Eqs.(\ref{mult})
and (\ref{exchange}) and encounters between particles do not lead to
any particular reaction event. 

In a separate publication \cite{tach2}, we proceed to show that this 
is also true for two-dimensional systems, while in 3D one may obtain 
a fluctuation-induced lower bound on the
decay function which,  in some range of parameters, is better
(higher) than predictions based on standard Smoluchowski approach, Eq.(\ref{3D}).

This paper is outlined as follows. In Section II we define the model and 
introduce basic notations for the general $d$-dimensional case. In Section III
we derive a general upper bound on the global decay function, 
while section IV presents the derivation of the lower bound. 
Next, in Sections V and VI, focusing on a 1D case, we evaluate the bounds on
 the global decay functions eplicitly for exchange-mediated and multipolar transfer, respectively, 
and demonstrate that they coincide in the asymptotic
limit $t \to \infty$ defining in such a way an asymptotically exact result. Finally, 
in Section VII we conclude with a brief recapitulation of our results and an outlook for future work.

\section{Model and basic equations}

Consider a $d$-dimensional spherical volume $V$ containing a single
excited donor molecule, which is initially located at the origin,
and $K$ acceptor molecules, placed at random positions. Suppose that
both donor and acceptors perform conventional diffusive motion with
diffusion coefficients $D_D$ and $D_A$, respectively. Let the
instantaneous positions of the donor and of the acceptors be denoted
by the ($d$-dimensional) vectors ${\bf r}(t)$ and $ {\bf R}_k(t)$,
$k=1,2,\ldots,K$.

We will neglect here the backtransfer to the donor. This neglect is
well justified if the donor-acceptor energy difference is much
larger than $k_B T$, $T$ being temperature and $k_B$ the Boltzmann
constant. We also disregard here donor-specific radiative and
radiationless processes. These decay channels are independent of the
direct energy transfer and thus the overall donor decay function
factorizes into the product of the donor-specific decay law,
$\exp(-t/\tau_R)$, where $\tau_R$ is the rate of the donor-specific
decay, times the acceptor determined decay function. Thus we focus
here only on the non-radiative donor-acceptor transfer.

One assumes the acceptors to act independently, which means that
they contribute multiplicatively to the decay. This assumption is
well fulfilled when the density of acceptors is low. Under such an
assumption, the probability that the donor is still in excited state
at time $t$, for a given realization of its trajectory ${\bf r}(t)$
and given realizations of acceptors' trajectories $\{{\bf
R}_k(t)\}$, is given by
\begin{equation}
\label{1} P({\bf r}(t),\{{\bf R}_k(t)\}) = \prod_{k=1}^K
\exp\left[-\int_0^t W(\rho_k(t')) dt' \right],
\end{equation}
where $\rho_k$ denotes the separation distance between the donor and
$k$-th acceptor.

Experimentally measured property is the global decay function
averaged over all possible donor and acceptor trajectories
\begin{equation}
\label{2} P(t) = E_{0}^D \left\{\left< \prod_{k=1}^K E_{{\bf
R}_k(0)}^{A}\left\{\exp\left[-\int_0^t W(\rho_k(t')) dt'
\right]\right\} \right>_{{\bf R}_k(0)}\right\},
\end{equation}
where the symbol $E_0^D\{...\}$ denotes averaging with respect to
all possible donor's trajectories ${\bf r}(t)$; symbols $E_{{\bf
R}_k(0)}^{A}\left\{...\right\}$ denote averaging with respect to the
trajectories of the k-th acceptor, commencing its motion at position
${\bf R}_k(0)$, and finally, the angle brackets stand for the
averaging with respect to the distribution of the starting
positions. Note that presenting $P(t)$ in the form as in
Eq.(\ref{2}), we have already implicitly assumed that all acceptors
move independently of each other, which is again well-justified for sufficiently low 
acceptor concentrations.

After some straightforward calculations, we arrive at the following
thermodynamic-limit expression:
\begin{equation}
\label{3} P(t) = E_{0}^D\left\{\exp\Big[- n_A Q({\bf
r}(t);t)\Big]\right\}
\end{equation}
where $n_A$ is the mean concentration of acceptor molecules, ($n_A =
K/V$ when both $K,V \to \infty$), while $Q({\bf r}(t);t)$ is the
following functional of a given donor trajectory ${\bf r}(t)$:
\begin{equation}
\label{nn} Q({\bf r}(t);t) = \int d{\bf R}(0) E_{{\bf R}(0)}^{A}
\left\{1-\exp\left[-\int_0^t dt' W(|{\bf r}(t') - {\bf
R}(t')|)\right]\right\}
\end{equation}
In the latter equation, ${\bf R}(t)$ denotes  a given trajectory of
a single acceptor molecule and $E_{{\bf R}(0)}^A\left\{...\right\}$
denotes averaging over all possible trajectories ${\bf R}(t)$. Note
that straightforward averaging in Eqs.(\ref{3}) and (\ref{nn}) is a
non-tractable mathematical problem since averaging over acceptor
trajectories in Eq.(\ref{nn}) has to be taken first for a
\textit{given} realization of donor's trajectory and only after
doing it, one may perform averaging of the exponential in
Eq.(\ref{3}). Consequently,  a recourse has to be made to
approximations.

\section{Upper bound on the global decay function: Pascal principle}

A convenient for our purposes upper bound on the global decay
function stems from the so-called Pascal principle, which in our
terms can be formulated as follows: an excitation on an immobile
donor molecule survives longer than on a randomly moving one. In
other words, $P(t)$ in Eq.(\ref{3}) is bounded by
\begin{equation}
\label{4} P(t) \leq P_{u}(t),
\end{equation}
where $P_{u}(t)$ describes the decay of an immobile donor, fixed at
the origin, due to a concentration $n_A$ of \textit{diffusive}
acceptor molecules,
\begin{equation}
\label{target} P_{u}(t)= \exp\left[- n_A \int d{\bf R}(0) E_{{\bf
R}(0)}^A \left\{1-\exp\left[-\int_0^t dt' W(|{\bf
R}(t')|)\right]\right\}\right]
\end{equation}
The inequality in Eq.(\ref{4}) has been first conjectured in
Ref.\cite{bb} for contact trapping reactions and proven in Ref.\cite{bmb}
for one-dimensional systems. In Ref.\cite{m}, Eq.(\ref{4}) has been
proven for a rather general class of random walks on $d$-dimensional
lattices. 
We also remark that a similar statement has been proven
earlier in Ref.\cite{b} for the process of an excitation energy
migration via distance-dependent transfer rates on a disordered
array of immobile donor molecules and quenched by randomly placed
immobile acceptors. It was shown that the survival probability of an
excitation can be only decreased because of random motion not
correlated with acceptors' spatial distribution.
However, no rigorous proof of such a statement exists at present 
for diffusion-controlled long-range reactions
although it is intuitively clear that
the inequality in Eq.(\ref{4}) should hold in this case too. 
We thus assume, without
proof, that the inequality in Eq.(\ref{4}) is also valid for the model under study.

Next, applying Feynmann-Kac theorem \cite{fh,kac} one may show that
\begin{equation}
\label{pp} E_{{\bf R}(0)}^A \left\{\exp\left[-\int_0^t dt' W(|{\bf
R}(t')|)\right]\right\} = \int d{\bf R} G_t({\bf R}|{\bf R}(0)),
\end{equation}
$G_t({\bf R}|{\bf R}(0))$ being the Green's function solution of the
following Schr\"odinger equation:
\begin{eqnarray}
\label{mm} \frac{\partial }{\partial t} G_t({\bf R}|{\bf R}(0)) &=&
D_A \triangle_{{\bf R}} G_t({\bf R}|{\bf R}(0)) - W(|{\bf R}|)
G_t({\bf R}|{\bf R}(0)), \nonumber\\ G_{t=0}({\bf R}|{\bf R}(0)) &=&
\delta({\bf R} - {\bf R}(0)),
\end{eqnarray}
where $\triangle_{{\bf R}}$ is a $d$-dimensional Laplace operator.

Note that Eqs.(\ref{mm}) presumes that donor and acceptors are
point-like, non-interacting particles. In reality, they possess
hard-cores and can not approach each other at distance less than
$a$, equal to the sum of donor and acceptor radii. This means that
Eqs.(\ref{mm}) are to be complemented  by a reflective boundary
condition at $|{\bf R}| = a$ \cite{bur}.

Taking advantage of Eqs.(\ref{pp}) and (\ref{mm}), we can formally
rewrite Eq.(\ref{target}) as
\begin{equation}
\label{target1} P_{u}(t)= \exp\left[n_A \int_0^t dt' \int d{\bf R}
\; \frac{\partial G_{t'}({\bf R})}{\partial t'}\right], \;G_{t}({\bf
R}) = \int d{\bf R}(0) G_{t}({\bf R}|{\bf R}(0)).
\end{equation}
Assuming next that $G_{t}({\bf R})$ is independent of angular
variables such that $G_{t}({\bf R}) = G_{t}(r)$, where $r = |{\bf
R}|$, we get the following compact expression:
\begin{eqnarray}
\label{target2} P_{u}(t)= \exp\left[ - n_A \int^t_0 dt' k_{u}(t')
\right],
\end{eqnarray}
in which equation $k_u(t)$ is determined by
\begin{equation}
\label{target3} k_{u}(t) =  d \; V_d \int^{\infty}_a r^{d-1} W(r)
G_{t}(r),
\end{equation}
and   $G_t(r)$ obeys
\begin{eqnarray}
\label{q} \frac{\partial G_t(r)}{\partial t} &=& D_A
\left(\frac{\partial^2 G_t(r)}{\partial r^2} + \frac{d-1}{r}
\frac{\partial
G_t(r)}{\partial r}\right) - W(r) G_t(r), \nonumber\\
G_{t=0} &=& 1; \;\;\; G_t(r \to \infty) = 1, \;\;\; \left.
\frac{\partial G_t(r)}{\partial r} \right|_{r = a} = 0.
\end{eqnarray}
Equations (\ref{target2}), (\ref{target3}) and (\ref{q}) thus define
the upper bound on the global decay function $P(t)$ in systems with
diffusive donor and acceptors.

\section{Lower bound on the global decay function.}

We turn now to the derivation of a lower bound on $P(t)$ in
Eq.(\ref{3}). Following Ref.\cite{bb} (see also Ref.\cite{rk}), we make
the
following steps:\\
(i) suppose that for a given initial placement of acceptors, a closest to the origin acceptor appears at distance $l$. Thus,
 a notional spherical volume $V_l$
of radius $l$, centered
at the origin, is initially completely devoid of acceptors.\\
(ii) performing averaging over donor's trajectories $\{{\bf
r}(t)\}$, we consider only such trajectories which never leave $V_l$
up to time moment $t$. Since $Q({\bf r}(t);t)$ in Eq.(\ref{nn}) is
always positive definite for any particular realization ${\bf
r}(t)$, such a constraint naturally leads to a lower bound on
$P(t)$, i.e.
\begin{equation}
E_{0}^D\left\{\exp\Big[- n_A Q({\bf r}(t);t)\Big]\right\} \geq E_{0,
{\bf r}(t) \in V_l}^D\left\{\exp\Big[- n_A Q({\bf
r}(t);t)\Big]\right\},
\end{equation}
where $E_{0, {\bf r}(t) \in V_l}^D\left\{...\right\}$ denotes
averaging over a subset of all possible donor's trajectories such
that they do not leave $V_l$ during time $t$.\\
(iii) considering the term responsible for long-range transfer,
$Q({\bf r}(t);t)$, we suppose that the donor is always located on the
surface of $V_l$ at position \textit{closest} to the instantaneous
position of the acceptor. Since $W(\rho)$ is a strictly decreasing
function of $\rho$,  for any $r(t) \in V_l$, one has $W(|{\bf R}(t)|
- l) \geq W(|{\bf r}(t) -{\bf R}(t)|)$ and hence, $Q({\bf r}(t);t)$
can be majorized by
\begin{equation}
\label{8} Q({\bf r}(t);t) \leq Q(l;t) = \int d{\bf R}(0) E_{{\bf
R}(0)}^{A} \left\{1-\exp\left[-\int_0^t dt' W(|{\bf R}(t')| -
l)\right]\right\}
\end{equation}

Note now that the right-hand-side of the inequality in Eq.(\ref{8})
is \textit{independent} of the donor's trajectories.

Consequently, collecting (i) to (iii), 
we arrive at the following \textit{lower} bound on the global decay
function
\begin{equation}
\label{zzu}
P(t) \geq P_{\rm void}(l)\; \times  E_{0, {\bf r}(t) \in
V_l}^D\left\{1\right\} \; \times \exp\left[- n_A R(l;t)\right].
\end{equation}
In this equation $P_{void}(l)$ is the probability of having an acceptor-free
spherical void of radius $l$. For random intial placement of acceptors, one has
\begin{equation}
\label{void} P_{\rm void}(l) \sim \exp\Big[- n_A V_d \; l^d\Big].
\end{equation}
Further on, in Eq.(\ref{zzu}) the symbol
$E_{0, {\bf r}(t) \in V_l}^D\left\{1\right\}$ denotes the measure of such
donor's trajectories, which commence at the origin and
never leave $V_l$ during time $t$; at sufficiently
large times, $E_{0, {\bf r}(t) \in V_l}^D\left\{1\right\}$ is given by
\begin{equation}
\label{measure} E_{0, {\bf r}(t) \in V_l}^D\left\{1\right\} \sim
\exp\left[ - z_d^2 \frac{D_D t}{l^2}\right],
\end{equation}
$z_d$ being the first zero of the Bessel function $J_{(d-2)/2}(x)$.

Combining the expressions in Eqs.(\ref{void}) and (\ref{measure}),
and assuming spherical symmetry, we finally obtain
\begin{equation}
\label{lower} P(t) \geq P_l(t) = \exp\left[ - n_A V_d l^d - z_d^2
\frac{D_D t}{l^2} - n_A  \int^t_0 dt' k_l(t')\right].
\end{equation}
In the latter equation,
\begin{equation}
\label{const}
 k_l(t) = d \; V_d
\int_{l+a}^{\infty} r^{d - 1} W(r - l) \tilde{G}_t(r) dr,
\end{equation}
while $\tilde{G}_{t}(r)$  is the solution of
\begin{eqnarray}
\label{qg} \frac{\partial \tilde{G}_t(r)}{\partial t} &=& D_A
\left(\frac{\partial^2 \tilde{G}_t(r)}{\partial r^2} + \frac{d-1}{r}
\frac{\partial
\tilde{G}_t(r)}{\partial r}\right) - W(r-l) \tilde{G}_t(r), \nonumber\\
\tilde{G}_{t=0}(r) &=& 1; \;\;\; \tilde{G}_t(r \to \infty) = 1,
\end{eqnarray}
subject, in virtue of condition (iii), to a reflection boundary
condition imposed at $r = l + a $:
\begin{equation} \label{rb} \left. \frac{\partial
\tilde{G}_t(r)}{\partial r} \right|_{r = l+ a} = 0.
\end{equation}
Equations (\ref{lower}),(\ref{const}),(\ref{qg}) and (\ref{rb})
define a family of lower bounds on the global decay function in
systems with diffusive donor and acceptors, dependent on the radius
$l$ of the notional volume $V_l$ encircling the donor and devoid of
acceptors. 

To get the optimal lower bound, we will have, in the
usual fashion, to maximize the result with respect to $l$. Below we
consider lower and upper bounds on the global decay function in
one-dimensional systems with long-range transfer
(Eqs.(\ref{mult}) and (\ref{exchange})) between diffusive donor and
diffusive acceptors. Corresponding results for two- and three-dimensional systems will be presented elsewhere \cite{tach2}.

\section{One-dimensional systems: Exchange-mediated transfer.}

\subsection{Upper bound.}

Consider first the derivation of \textit{an upper bound} in
one-dimensional systems with a transfer mediated  by exchange. Here,
Laplace-transformed with respect to time variable $t$, solution of
Eqs.(\ref{qg}) and (\ref{rb}) reads:
\begin{eqnarray} \label{sol} G_{\lambda}(r) &=& \int^{\infty}_0 dt
\exp[- \lambda t] G_t(r) = C_1 I_{\chi}(x) + C_2
K_{\chi}(x) + \nonumber\\
&+&\frac{\chi \Gamma(1-\frac{\chi}{2})}{\lambda}
\left(\frac{x}{2}\right)^{\frac{\chi}{2}} \int^{1}_0
I_{-\frac{\chi}{2}}\left(x \xi\right) \xi^{1+\frac{\chi}{2}}
\left(1-\xi^2\right)^{\frac{\chi}{2}-1} d\xi,
\end{eqnarray}
where $K_{\chi}(x)$ and $I_{\chi}(x)$ are modified Bessel functions,
the integral term in the second-line is a particular solution
(Lommel function) and
\begin{eqnarray}
x= \omega \exp[-\gamma \frac{r}{2}], \;\;\; x_0 = \omega
\exp[-\gamma \frac{a}{2}], \;\;\; \omega = \frac{2}{\gamma}
\sqrt{\frac{\alpha_e}{D_A}}, \;\;\; \text{and} \;\;\;
\chi=\frac{2}{\gamma} \sqrt{\frac{\lambda}{D_A}}.
\end{eqnarray}
Now, note that as $r \to \infty$, $x \to 0$, $I_{\chi}(x) \to 0$,
the last term on the rhs of Eq.(\ref{sol}) tends to $1/\lambda$,
while $K_{\chi}(x)$ diverges. Hence, we set $C_2 =0$. Further, we
get that the reflective boundary condition at the closest approach
distance is fulfilled when
\begin{equation}
\label{C1} C_1 = - \frac{2 \chi \Gamma(1-\frac{\chi}{2})}{\lambda
\left(I_{\chi-1}(x_0) + I_{\chi+1}(x_0)\right)}
\left(\frac{x_0}{2}\right)^{\frac{\chi}{2}} \int^{1}_0
I_{1-\frac{\chi}{2}}\left(x_0 \xi\right) \xi^{2+\frac{\chi}{2}}
\left(1-\xi^2\right)^{\frac{\chi}{2}-1} d\xi
\end{equation}
Plugging Eqs.(\ref{sol}) and (\ref{C1}) into Eq.(\ref{target3}) and
performing intergation, we find that the Laplace-transformed reaction
constant $k_u(\lambda)$ is given by
\begin{eqnarray}
\label{kk} k_u(\lambda) &=& \frac{D_A \gamma x_0^2}{\lambda} \Big[
\;_2F_3\left(1,1;2,1-\frac{\chi}{2},1+\frac{\chi}{2},\frac{x_0^2}{4}\right)
- \frac{2}{(1+\frac{\chi}{2}) (1-\frac{\chi^2}{4}) \Gamma(1+\chi)}
\times \nonumber\\
&\times& \left(\frac{x_0}{2}\right)^{1+\chi} \frac{\;_1F_2\left(2;
2- \frac{\chi}{2},2+\frac{\chi}{2};\frac{x_0^2}{4}\right)
\;_1F_2\left(1+\frac{\chi}{2}; 2+
\frac{\chi}{2},1+\chi;\frac{x_0^2}{4}\right)
}{I_{\chi-1}\left(x_0\right) + I_{\chi+1}\left(x_0\right)}\Big],
\end{eqnarray}
where $_pF_q$ denote generalized hypergeometric functions.

Leading small-$\lambda$ (large-$t$) asymptotic behavior of
$k_u(\lambda)$ in Eq.(\ref{kk}) follows
\begin{equation}
\label{mmm} k_u(\lambda) \sim 2 \sqrt{\frac{D_A}{\lambda}}
\frac{1}{1 + \sqrt{T_e \lambda}},
\end{equation}
where
\begin{equation}
\label{te} T_e = \left(\frac{K_1\left(x_0\right) + \left(1/2 - C +
\ln\left(2/x_0\right)\right)
I_1\left(x_0\right)}{I_1\left(x_0\right)}\right)^2 \frac{4}{\gamma^2
D_A},
\end{equation}
$C \approx 0.577$ being the Euler constant.

This yields, in $t$-domain, the following asymptotical behavior
\begin{equation}
\label{contu} \int^t_0 dt' k_u(t') = 4 \sqrt{\frac{D_A
t}{\pi}}\left(1 -
 \sqrt{\frac{\pi T_e}{4 t}} +
O\left(\frac{1}{t}\right)\right)
\end{equation}
Consequently, in 1d systems with transfer mediated by exchange we
have the following upper bound on the global decay function:
\begin{equation}
\label{upper1dex} P(t) \leq \exp\left[ - 4 n_A \sqrt{\frac{D_A
t}{\pi}} + 2 n_A \sqrt{D_A T_e} +
O\left(\frac{1}{t^{1/2}}\right)\right]
\end{equation}
Before we proceed to the derivation of the lower bound, a few
comments are in order:\\
(a) first of all, we notice that the right-hand-side of
Eq.(\ref{upper1dex}) coincides with the solution of the so-called
target problem - probability that an immobile target survives, in
one dimension, up to time $t$ in presence of diffusive scavengers
which may "destroy" the target upon the first encounter with it
\cite{target}. Therefore, in one dimension, at sufficiently long
times the kinetic behavior of long-range transfer proceeds exactly
in the same way as for contact diffusion-limited target annihilation
reaction, despite the fact that here the boundary condition imposed
on the donor's surface is reflective and the deactivation of the
donor happens, at rate
$\alpha_e \exp(- \gamma r)$, at any donor-acceptor distance $r$.\\
(b) parameter $T_e$ in Eq.(\ref{te}) is the crossover time to the
asymptotic stage $\ln P_u(t) \sim - t^{1/2}$ for exchange-mediated
transfer in one-dimensional systems with immobile donor and mobile
acceptors. Note that $\gamma^2 D_A T_e$ is a non-monotonic function
of $x_0$. It is large $\sim 1/x_0^4$ when $x_0 \ll 1$ (i.e., when
$D_A$ is large), such that $T_e \sim \gamma^2 D_A/\alpha_e^2$. In
this case, one would first observe, for $0 < t < T_e$, an
intermediate asymptotical behavior $\ln P(t) \sim - \gamma \alpha
t$, which will then cross to the asymptotical behavior in
Eq.(\ref{upper1dex}). Next, note that $\gamma^2 D_A T_e$ is also
large when $x_0 \gg 1$, which happens when $D_A$ is small. Here,
$T_e \approx \ln^2(x_0/2)/\gamma^2 D_A$, i.e. $T_e$ is proportional
to the first inverse power of $D_A$ (with logarithmic corrections).
In this case, the asymptotic decay in Eq.(\ref{upper1dex}) succeeds
the static quenching decay in Eq.(\ref{exchange1}), which is valid
in progressively larger time
domain the closer $D_A$ is to zero.\\
(c) finally, we remark that despite the fact that the result in
Eq.(\ref{contu}) is independent of both $\alpha_e$ and $\gamma$,
which are the only parameters characterizing the transfer rate and
thus "represent" reaction, it does not mean that it can be simply
obtained by expanding $G(r) = \sum_{n=0}^{\infty} \alpha_e^n G_n(r)$
and considering the zeroth term only. In general, Eq.(\ref{contu}) is
essentially a non-perturbative result and can not be obtained using
a perturbative expansion of $G_t(r)$ in powers of $\alpha_e$,
unless, of course, one manages to sum the whole series. On the other
hand, Eq.(\ref{contu}) can be straightforwardly derived
approximating the transfer rate by a step-function ("square well"
approximation).

Indeed, suppose that $\gamma a < 1$ and consider separately solution
of Eqs.(\ref{qg}) and (\ref{rb}) for $a  \leq r \leq 1/\gamma$ and $
r \geq 1/\gamma$. In the first interval we approximate $\exp[-\gamma
r]$ by $\exp[-\gamma a]$, and find that the Laplace-transformed
solution of the Schr\"odinger equation which obeys the reflecting
boundary condition reads
\begin{equation}
G_{\lambda}^{(1)} = \frac{1}{\lambda+ \alpha_e \exp[-\gamma a]}  +
C_1 \cosh\left[\sqrt{\frac{\lambda+ \alpha_e \exp[-\gamma a]}{D_A}}
(r-a)\right]
\end{equation}
On the other hand, in the domain $r \geq 1/\gamma$, the transfer
term can be neglected, and we have
\begin{equation}
G_{\lambda}^{(2)}(r) = \frac{1}{\lambda}  + C_2 \exp\left[-
\sqrt{\frac{\lambda}{D_A}} \left(r - \frac{1}{\gamma}\right)\right]
\end{equation}
Since $G_{\lambda}(r)$ and its first derivative have to be
continuous functions at $r = 1/\gamma$, we have two complementary
equations which define the coefficients $C_1$ and $C_2$. Determining
these coefficients, we find that the leading small-$\lambda$
behavior of $k_u(\lambda)$ follows
\begin{eqnarray}
\label{kkq} k_{u}(\lambda) &=&  2 \alpha_e \exp\left[-\gamma a
\right] \int^{1/\gamma}_a G_{\lambda}^{(1)}(r) dr = 2
\sqrt{\frac{D_A}{\lambda}} \frac{1}{1 + \sqrt{T_e' \lambda}},
\end{eqnarray}
where
\begin{equation}
T_e'= \frac{\exp[a \gamma]}{\alpha_e}
\coth^2\left(\frac{x_0}{2}\right).
\end{equation}
Note that $k_u(\lambda)$ in Eq.(\ref{kkq}) has exactly the same form
as $k_u(\lambda)$ in Eq.(\ref{mmm}), which means that the "square
well" approximation captures well the leading behavior of the
effective reaction rate. The crossover time $T_e'$ has a different
form compared to the exact one, Eq.(\ref{te}); it exhibits, however,
quite a "correct" behavior in the case  $x_0 \ll 1$ (fast diffusion)
when $T_e' \sim \gamma^2 D_A/\alpha_e^2$.

\subsection{Lower bound.}
 
Consider now a lower bound on $P(t)$ for one-dimensional systems
with transfer mediated by exchange interactions. Laplace-transformed
solution of Eqs.(\ref{qg}) and (\ref{rb}) reads
\begin{eqnarray}
\label{soll} \tilde{G}_{\lambda}(r) = C_1 I_{\chi}(x e^{\frac{\gamma l}{2}}) + \frac{\chi \Gamma(1-\chi/2)}{\lambda}
\left(\frac{x e^{\frac{\gamma l}{2}}}{2}\right)^{\frac{\chi}{2}}
\int^{1}_0 I_{-\frac{\chi}{2}}\left(x \xi e^{\frac{\gamma
l}{2}}\right) \xi^{1+\frac{\chi}{2}}
\left(1-\xi^2\right)^{\frac{\chi}{2}-1} d\xi,
\end{eqnarray}
where $C_1$ is given by Eq.(\ref{C1}). Plugging the expression in
Eq.(\ref{soll}) into Eq.(\ref{const}) and performing integration, we
find that $k_l(\lambda)$ obeys
\begin{equation}
k_l(\lambda) \equiv k_u(\lambda),
\end{equation}
where $k_u(\lambda)$ is determined by Eq.(\ref{kk}). Consequently,
the lower bound on $P(t)$, Eq.(\ref{lower}), at sufficiently long
times attains the following form:
\begin{equation}
\label{lower1} P_l(t) \sim \exp\left[ - 2 n_A l - \pi^2 \frac{D_D
t}{l^2} - 4 n_A  \sqrt{\frac{D_A t}{\pi}}\right]
\end{equation}

As we have already mentioned, the result in Eq.(\ref{lower1})
represent rather a family of lower bounds dependent on parameter $l$
-  radius of a notional volume initially devoid of acceptors. The
"best" lower bound thus would be the highest one. Optimizing
Eq.(\ref{lower1}) with respect to $l$, we find that the highest
lower bound is achieved when $l = (\pi^2 D_D t/n_A)^{1/3}$, and is
given by
\begin{equation}
\label{lmax}
P_{l,max}(t) \sim \exp\left[- 4 n_A  \sqrt{\frac{D_A
t}{\pi}} - 3 n_A^{2/3}\left(\pi^2 D_D t\right)^{1/3} \right]
\end{equation}
On comparing the asymptotic behavior predicted by the maximal lower
bound in Eq.(\ref{lmax}) against the upper bound in
Eq.(\ref{upper1dex}) we notice that both bounds converge
asymptotically to give an exact result in Eq.(\ref{1d1d}).

\section{One-dimensional systems: Multipolar transfer.}

\subsection{Upper bound.}

Consider now, within the "square well" approximation, an upper bound
in case of multipolar transfer in Eq.(\ref{mult}). Approximating the
actual transfer rate $W(r)$ in Eq.(\ref{mult}) by a step-function
\begin{equation}
W(r) = \left\{\begin{array}{ll}
\alpha_m (r_0/a)^n,    \mbox{$a \leq r \leq r_0$, } \nonumber\\
0,    \;\;\; \mbox{$r \geq r_0$}
\end{array}
\right.
\end{equation}
we find that in the interval $a \leq r \leq r_0$ Laplace-transformed
solution of Eq.(\ref{q}) obeys
\begin{equation}
\label{ggg} G_{\lambda}^{(1)} = \frac{1}{\lambda+ \alpha_m
(r_0/a)^n}  + C_1 \cosh\left[\sqrt{\frac{\lambda+ \alpha_m
(r_0/a)^n}{D_A}} (r-a)\right],
\end{equation}
while in the domain $r \geq r_0$ it follows
\begin{equation}
G_{\lambda}^{(2)}(r) = \frac{1}{\lambda}  + C_2 \exp\left[-
\sqrt{\frac{\lambda}{D_A}} \left(r - r_0\right)\right].
\end{equation}
Constants $C_1$ and $C_2$ are to be chosen in such a way that both
$G_{\lambda}(r)$ and its first derivative are continuous functions
at $r = r_0$.

Determining these constants, plugging Eq.(\ref{ggg}) into
Eq.(\ref{target3}) and performing integration, we find that the
Laplace-transformed $k_u(t)$ is given by
\begin{equation}
k_u(\lambda) = 2 \sqrt{\frac{D_A}{\lambda}} \left(1 +
\sqrt{\frac{\lambda a^n}{\alpha_m r_0^n}} \coth\left(\sqrt{\frac{\alpha_m (r_0/a)^n}{D_A}}(r_0 -
a)\right)\right)^{-1} + \frac{2
\alpha_m (r_0/a)^n (r_0 - a)}{\lambda + \alpha_m (r_0/a)^n}
\end{equation}
This yields, in $t$-domain,
\begin{equation}
\int^t_0 k_u(t') dt' = 4 \sqrt{\frac{D_A t}{\pi}}\left(1 -
 \sqrt{\frac{\pi a^n}{4 \alpha_m r_0^n \, t}} \coth\left(\sqrt{\frac{\alpha_m (r_0/a)^n}{D_A}}(r_0 -
a)\right) +
O\left(\frac{1}{t}\right)\right).
\end{equation}
Consequently, the global decay function $P(t)$ in one-dimensional
systems with diffusive donor and acceptors interacting via
multipolar transfer rate in Eq.(\ref{mult}) is bounded from above by
\begin{equation}
\label{llmax}
P(t) \leq \exp\left[ - 4 n_A \sqrt{\frac{D_A t}{\pi}} + 2 n_A
\sqrt{\frac{D_A a^n}{\alpha_m r_0^n}} \coth\left(\sqrt{\frac{\alpha_m
(r_0/a)^n}{D_A}}(r_0 - a)\right) \right].
\end{equation}

\subsection{Lower bound.}

Turning next to evaluation of the lower bound on $P(t)$ we introduce
parameter $\delta > a$, and approximate the actual transfer rate by
a step-function of the form
\begin{equation}
W(r) = \left\{\begin{array}{ll}
\alpha_m (r_0/a)^n,    \mbox{$l + a \leq r \leq l + \delta$, } \nonumber\\
0,    \;\;\; \mbox{$r \geq \delta$}
\end{array}
\right.
\end{equation}
Approximate solution of Eqs.(\ref{qg}) in the interval $l + a \leq r
\leq l + \delta$ has the form
\begin{equation}
\label{gggg} \tilde{G}_{\lambda}^{(1)} = \frac{1}{\lambda+ \alpha_m
(r_0/a)^n}  + C_1 \cosh\left[\sqrt{\frac{\lambda+ \alpha_m
(r_0/a)^n}{D_A}} (r-l-a)\right],
\end{equation}
while in the domain $r \geq l + \delta$ it is given by
\begin{equation}
\tilde{G}_{\lambda}^{(2)}(r) = \frac{1}{\lambda} + C_2 \exp\left[-
\sqrt{\frac{\lambda}{D_A}} \left(r - l - \delta\right)\right].
\end{equation}
Again, requiring continuity of $\tilde{G}_{\lambda}(r)$ and of its first
derivative at $r = l + \delta$, we determine $C_1$ and $C_2$, which
yields, after straightforward calculations, the following expression
\begin{equation}
k_l(\lambda) = 2 \sqrt{\frac{D_A}{\lambda}} \left(1 +
\sqrt{\frac{\lambda a^n}{\alpha_m r_0^n}} \coth\left(\sqrt{\frac{\alpha_m
(r_0/a)^n}{D_A}}(\delta - a)\right)\right)^{-1} + \frac{2 \alpha_m (r_0/a)^n (\delta -
a)}{\lambda + \alpha_m (r_0/a)^n},
\end{equation}
We find then that in the $t$-domain, the leading behavior of
$\int^t_0 k_l(t') dt'$ is given by
\begin{equation}
\int^t_0 k_l(t') dt' = 4 \sqrt{\frac{D_A t}{\pi}} - 2
\sqrt{\frac{D_A a^n}{\alpha_m r_0^n}} \coth\left(\sqrt{\frac{\alpha_m
(r_0/a)^n}{D_A}}(\delta - a)\right)+
O\left(\frac{1}{t^{1/2}}\right)
\end{equation}
Consequently, 
an optimized lower bound on $P(t)$ reads
\begin{equation}
\label{lmax2} P(t) \geq \exp\left[- 4 n_A  \sqrt{\frac{D_A t}{\pi}}
- 3 n_A^{2/3}\left(\pi^2 D_D t\right)^{1/3} + 2 n_A \sqrt{\frac{D_A a^n
}{\alpha_m r_0^n}} \coth\left(\sqrt{\frac{\alpha_m (r_0/a)^n}{D_A}}(\delta -
a)\right) \right]
\end{equation}
On comparing the results in Eqs.(\ref{llmax}) and (\ref{lmax2}), we notice that again both bounds converge as $t \to \infty$
determining exact asymptotic decay of the excited donor,
Eq.(\ref{1d1d}).

\section{Conclusions}

To conclude, we have 
studied analytically 
direct energy transfer between diffusive excited donor and diffusive unexcited
acceptors mediated by multipolar  or exchange
interactions. 
Extending
a non-perturbative approach by Bray and Blythe
\cite{bb}
 (originally developed for
contact diffusion-controlled reactions)  
over the case of long-range
transfer, we have determined 
exactly
long-time asymptotics of the donor decay function in one-dimensional
systems. 
We have shown that the leading long-time behavior is independent of the diffusion constant $D_D$ 
of the donor molecule, and has exactly the same form as that descibing contact process. This finding is in aparent contradiction
with the results in Eqs.(\ref{3D}) and (\ref{low}).

We 
proceed to show elsewhere \cite{tach2} 
that 
also 
in two-dimensional systems the leading long-time behavior will be independent of  $D_D$, while 
in 3D a similar approach will give rise to
a fluctuation-induced lower bound on the
decay function 
which,  in some range of parameters, is better
(higher) than predictions based on standard Smoluchowski approach.

\section{Acknowledgments}

Research of GO is partially supported by Agence Nationale de la Recherche
(ANR) under grant ``DYOPTRI - Dynamique et Optimisation des Processus de
Transport Intermittents''. GO also acknowledges partial support and hospitality
of AIST-Tsukuba.


\begin{thebibliography}{10}

\bibitem{rice} S.A.Rice, {\it Diffusion-Limited Reactions}, in:
C.H.Bamford, C.F.H.Tipper, R.G.Compton (Eds), Comprehensive Chemical
Kinetics, Vol. {\bf 25} (elsevier, Amsterdam, 1985)

\bibitem{bur} A.I.Burshtein, Sov. Phys. Usp. {\bf 27}, 579 (1984)

\bibitem{graetzel} M.Graetzel, {\it Heterogeneous Photochemical Electron
Transfer}, (CRC Press, Boca Raton, FL, 1989)

\bibitem{klafter} J.Klafter and J.M.Drake (Eds.) {\it Molecular Dynamics in Restricted
Geometries}, (Wiley, New York, 1989)

\bibitem{bar} A.V.Barzykin, P.A.Frantsuzov, K.Seki and M.Tachiya,
Adv. Chem. Phys. {\bf 123}, 511 (2002)

\bibitem{f} T.F\"orster, Z. Naturforsh. Teil A {\bf 4}, 321 (1949)

\bibitem{d} D.L.Dexter, J. Chem. Phys. {\bf 21}, 836 (1953)

\bibitem{2} see, e.g., M.Inokuti and F.Hirayama, J. Chem. Phys. {\bf 43}, 1978
(1965); M.Tachiya and A.Mozumder, Chem. Phys. Lett. {\bf 28}, 87
(1974); A.Blumen and R.Silbey, J. Chem. Phys. {\bf 70}, 3707 (1979)

\bibitem{26} A.Blumen, J.Klafter and G.Zumofen, in {\it Optical Spectroscopy of
Glasses}, Ed. I.Zschokke, (Reidel, Dordrecht, 1986), p.199

\bibitem{lev} P. Levitz and J. M. Drake, Phys. Rev. Lett. {\bf 58}, 686 (1987)

\bibitem{lev2} J. M. Drake et al.,
Phys. Rev. Lett. {\bf 61}, 865 (1988)


\bibitem{microhet}  A.V.Barzykin, K.Seki and M.Tachiya, Adv. Colloid
Interface Sci.  {\bf 89}, 47 (2001).

\bibitem{polexp} O. Pekcan, M.A.Winnik and M.D.Croucher, Phys. Rev. Lett. {\bf 61}, 641 (1988);
A.Moeglich, K.Joder and T.Kiefhaber, PNAS {\bf 103}, 12394
(2006)

\bibitem{polteor} A. K. Roy and A. Blumen,
J. Chem. Phys. {\bf 91}, 4353 (1989); S.F.Burlatsky, G.S.Oshanin,
and A.V.Mogutov, Phys. Rev. Lett. {\bf 65}, 3205 (1990); G. Oshanin,
A. Blumen, M. Moreau, and S. F. Burlatsky, J. Chem. Phys. {\bf 103},
9864 (1995);
 J.H.Kim and S.Lee, J. Chem. Phys.{\bf
 119}, 11957
(2003); R.Reigada and I.M.Sokolov, Macromolecules {\bf 38}, 3504
(2005)

\bibitem{b} S.F.Burlatsky, G.Oshanin and A.A.Ovchinnikov,
Phys. Lett. A {\bf 139}, 241 (1989)

\bibitem{fed} S.G.Fedorenko, A.I.Burshtein and A.A.Kipriyanov, Phys. Rev. B {\bf 48}, 7020 (1993)

\bibitem{yt} see, e.g., M.Yokota and O.Tanimoto, J. Phys. Soc. Jpn {\bf 22},
779 (1967); I.Z.Steinberg and E.Katchalski, J. Chem. Phys. {\bf 48},
404 (1968); M.J.Pilling and S.A.Rice, J. Chem. Soc. Faraday Trans.
{\bf 72}, 792 (1976); U.G\"osele, M.Hauser, U.K.A.Klein and R.Frey,
Chem. Phys. Lett. {\bf 34}, 519 (1975); K.Allinger and A.Blumen, J.
Chem. Phys. {\bf 72}, 4608 (1980); B.Sipp and R.Voltz, J. Chem.
Phys. {\bf 79}, 434 (1983)

\bibitem{pdg} P.-G.de Gennes, J. Phys. Chem. Solids {\bf 7}, 345
(1958)

\bibitem{seki-san} K.Seki, A.I.Shushin, M.Wojcik and M.Tachiya, J.
Phys. Cond. Mat. {\bf 19}, 065117 (2007)

\bibitem{pastur} L.A.Pastur, Theor. Math. Phys. {\bf 32}, 88 (1977)

\bibitem{bb} A.J.Bray and R.A.Blythe, Phys. Rev. Lett. {\bf 89},
150601 (2002); Phys. Rev. E {\bf 67} 041101 (2003)

\bibitem{bmb} A.J.Bray, S.N.Majumdar and R.A.Blythe, Phys. Rev. E {\bf 67},
060102R (2003)

\bibitem{subdiff} G.Oshanin, O.B\'enichou, M.Coppey, and M.Moreau,
Phys. Rev. E {\bf 66}, 060101 (2002); S.B.Yuste and K.Lindenberg,
Phys. Rev. E {\bf 72}, 061103 (2005); S.B.Yuste, J.J.Ruiz-Lorenzo,
and K.Lindenberg, Phys. Rev. E {\bf 74}, 046119 (2006); S.B.Yuste, G.Oshanin, K.Lindenberg, O.B\'enichou and 
J.Klafter, Survival probability of a particle in a sea of mobile traps: A tale of tails,
arXiv:0805.2920, appearing in PRE

\bibitem{tach2} G.Oshanin and M.Tachiya, in preparation

\bibitem{m} M.Moreau, G.Oshanin, O.B\'enichou and M.Coppey, Phys. Rev. E {\bf 67},
045104R (2003); Phys. Rev. E {\bf 69}, 046101 (2004)

\bibitem{rk} S.Redner and K.Kang,
J. Phys. A {\bf 17}, L451 (1984)

\bibitem{fh} R.P.Feynmann and A.Hibbs, {\it Quantum mechanics and
path integrals}, (McGraw-Hill, New York, 1965)

\bibitem{kac} M.Kac, {\it Probability and related topics in physical
sciences}, (Interscience, London, 1959)

\bibitem{target} M.Tachiya, Radiat. Phys. Chem. {\bf 21}, 167
(1983); A.Blumen, G.Zumofen and J.Klafter, Phys. Rev. B {\bf 30},
5379 (1984); A.Szabo, R.Zwanzig and N.Agmon, Phys. Rev. Lett. {\bf
61}, 2496 (1988); S.F.Burlatsky, M.Moreau, G.Oshanin and A.Blumen,
Phys. Rev. Lett. {\bf 75}, 585 (1995); D.P.Bhatia, M.A.Prasad and
D.Arora, {\bf 75}, 586 (1995); O.B\'enichou, M.Moreau and G.Oshanin,
Phys. Rev. E {\bf 61}, 3388 (2000).


\end{thebibliography}
\end{document}